# Exact quantization of the Milson potential
# via Romanovski-Routh polynomials

## G. Natanson

ai-solutions Inc.

2232 Blue Valley Dr.
Silver Spring MD 20904
U.S.A.
greg_natanson@yahoo.com

The paper re-examines Milson's analysis of the rational canonical Sturm-Liouville equation (RCSLE) with two complex conjugated regular singular points $-i$ and $+i$ by taking advantage of Stevenson's complex linear-fraction transformation $\xi_S(\eta)$ of the real variable $\eta$ ( $-\infty < \eta < +\infty$). It was explicitly shown that Stevenson's hypergeometric polynomials in a complex argument $\xi_S$ are nothing but Romanovski-Routh polynomials converted from $\eta$ to $\xi_S$. The use of Stevenson's mathematical arguments unambiguously confirmed 'exact solvability' of the Milson potential.

It was revealed that the Milson potential has two branches referred to as 'inside' and 'outside' depending on positions of zeros of the so-called 'tangent polynomial' (TP) relative to the unit circle. The two intersect along the shape-invariant Gendenshtein (Scarf II) potential. The remarkable feature of the RCSLE associated with the inner branch of the Milson potential (as well as its shape-invariant limit) is that it has two sequences of nodeless almost-everywhere holomorphic (AEH) solutions which can be used as factorization functions (FFs) for constructing new quantized-by-polynomials potentials. In case of the Gendenshtein potential complex-conjugated characteristic exponents (ChExps) at finite singular points of the given RCSLE become energy independent so that each polynomial sequence turns into a finite set of orthogonal polynomials. This confirms Quesne's conjecture [*J. Math. Phys.* **54** 122103 (2013)] that the 'Case III' polynomials discovered by her can be used for constructing orthogonal polynomials of novel type.

## 1. Introduction

About forty years ago the author [1] found a 6-parameter family of rational potentials exactly solvable in terms of hypergeometric functions of a *real* variable z. We referred to them below as 'r-Gauss Reference' (*r*-GRef) potentials to distinguish the latter from a family of '*i*-Gauss Reference' (*i*- GRef) potentials discovered by Milson [2]. As pointed to in [2] (see also a more recent review of Dereziński and Wrochna [3]), the conversion formulas adopted by the author from Bose's work [4] are nothing but the Liouville transformation [5-9] of the rational Sturm-Liouville equation (SLE) written in the 'normal' [6, 10, 11] or 'canonical' [2] form. In our papers we will refer to this equation as rational canonical SLE (RCSLE).

During his earlier studies of solvable potentials the author did search for possible extensions of the developed formalism but unfortunately overlooked a very important work by Stevenson [12] who explicitly expressed bound wave functions for the trigonometric Schrödinger (t-Schrödinger) potential [13] in terms of hypergeometric polynomials in a complex argument. As demonstrated in this paper, the solution found by Stevenson provides the accurate mathematic foundation for quantization of the RCSLE with two regular complex-conjugated singular points.

Though the RCSLE of our interested has been already studied by Milson [2] in great details, his analysis was done again with no reference to Stevenson's work. The most important difference, compared with this paper, is that Milson only proved that the potential in question is *'quasi-exactly'* [14-18] solvable. On other hand, the use of Stevenson's technique [12] allows us to prove that the generic '*i*-GRef' potential is *exactly* quantized by polynomials, which is also true for its two 'shape-invariant' limits: for the aforementioned t-Schrödinger potential (the 'trigonometric Rosen-Morse' potential in terms of Cooper et al [19, 20]) -- as it was proven by Stevenson himself -- and for the so-called [19, 20] 'Scarf II' potential referred to below (in following [21]) as the Gendenshtein [22] potential.

Regrettably, Stevenson's work [12] was overlooked not only by us [1] but by most researches in the field and for many years exact solutions of the RCSLE with two complex-conjugated singular points were simply obtained by expressing the appropriate bound eigenfunctions in terms of the complexified Jacobi polynomials. To our knowledge, it was Bagrov et al. [23], who first applied this extension of classical Jacobi polynomials [24] to bound states in the t-Schrödinger potential, though without providing any proof that the found solutions are real functions and that no other bound solutions are possible.

In this respect a crucial step forward was made in the work of Dabrowska, Khare, and Sukhatme [25] who addressed the problem from an entirely fresh view. Namely, they used sequential Darboux transformations [26, 27] to express eigenfunctions of bound states in the Gendenshtein potential via a sequence of complexified polynomials in an imaginary argument. It was shown that the resultant recurrent relations coincide with those for classical Jacobi polynomials [28], provided that the coefficients in the latter relations were extended over the field of complex numbers. Their approach automatically guaranteed that the n[th] bound state was described by the polynomial of order n with n real roots inside the quantization interval. Though Dabrowska et al failed to prove that they deal with real polynomials, this was necessarily assured by the way in which these polynomials were introduced.

Approximately at the same time but in a completely different field Askey [29] introduced Jacobi polynomials with complex-conjugated indexes and proved that they have an orthogonal subset when restricted to the imaginary axis. He also explicitly related the latter to Romanovsky's classical study [30] on finite subsets of orthogonal polynomials. However, this relationship was overlooked in the quantum-mechanical literature [31-34] for nearly two decades until it was thoroughly elaborated in more recent works of Kirchbach et al [35-37].

Before proceeding with this concise historical survey let us briefly summarize relevant results of four publications [31-34] mentioned above due to their significant impact on future developments in the field. First, the expressions for bound eigenfunctions in terms of Jacobi polynomials with complex-conjugated indexes [25] were promptly extended by Levai [31] to the t-Schrödinger potential (which was to a certain extent a re-discovery of the exact solutions reported earlier by Bagrov et al. [23]). At the same time Bagrov and Gitman [32] published an updated version of their book where explicit expressions for bound eigenfunctions were also presented for the Gendenshtein potential. Again the new book contained no relevant references and no proof was presented that the polynomials in question have real coefficients.

It was Jafarizadeh and Fakri [33] who traced the polynomial solutions constructed in [25] to the generating function for Romanovski/pseudo-Jacobi polynomials [30] (using Lesky's terms [38][x)]) which was re-discovered by these authors in previous publications [40, 41]. Subsequently Jafarizadeh and Fakri's results [33] were elaborated by Cotfas [34] based on the Nikiforov-Uvarov formalism [42] for generating orthogonal polynomials of hypergeometric type. Independently, Koepf and Masjed-Jamei [43] presented a systematic analysis of six sets of orthogonal polynomial associated with differential equation of hypergeometric type directly referring the one of our current interest to Romanovsky's work [30].

Ultimately Kirchbach et al in the already cited works [35-37] approached the quantization of the shape-invariant $i$-GRef potentials within historically precise framework by explicitly relating the appropriate normalizable eigenfunctions to Romanovski/pseudo-Jacobi polynomials (simply referred to in their papers as Romanovski polynomials). It should be stressed in this connection that the mathematical arguments presented by these authors again only indicated that both t-Schrödinger and Gendenshtein potentials are quasi-exactly solvable in terms of the Romanovski polynomials in question. One can argue that exact solvability of both potentials directly follows from their shape invariance. However, contrary to the wide-spread view, the shape invariance on its own only guarantees that the given potential is quasi-exactly solvable. As illustrated in Appendix A using the Gendenshtein potential as an example, the proof of its exact solvability based solely on the shape invariance requires some additional not-easy-to-verify assumptions.

The cited works of Kirchbach et al drew author's attention both to a finite set of orthogonal polynomials discovered by Romanovsky as well as to Routh's little known study [44] on all possible polynomial solutions of the hypergeometric-type differential

________________

[x)] In the mathematical literature these polynomials are also referred to as ''twisted Jacobi polynomial systems' [39].

equation with complex-conjugated singular points. To our knowledge, the mentioned reference to Routh' work was taken by Kirchbach et al from Ismail's monograph [45] who, in our judgment, misinterpreted Routh' results. As explained in Appendix B the real significance of Routh' work comes from his proof that the mentioned equation of hypergeometric type has an infinite set of polynomial solutions referred to below as Routh polynomials. Contrary to Ismail's statement, we could found no mention of a Rodrigues formula in [44]. An analysis of (20.1.1)-(20.1.3) in [45] shows that one of three sets of orthogonal polynomials discovered by Romanovsky [30] is indeed formed by Routh polynomials but the question of orthogonality was simply out of scope of Routh' paper. Unfortunately misinterpretation of Routh'work in [45] led many authors [37, 46-50] to an erroneous conclusion that Romanovsky re-discovered Routh polynomials.

Keeping in mind that the Romanovski polynomials in question form an orthogonal subset of an infinite set of Routh polynomials, it seems more appropriate to refer to them as 'Romanovski-Routh' polynomials, by analogy with the terms 'Romanovski-Bessel' and 'Romanovski-Jacobi' polynomials adopted in [43] from Lesky's paper [38]. (To be fully consistent with Lesky's terminology, we changed Askey's transliteration of Romanovsky's name to its French version 'Romanovski'.) A very close term 'Routh-Romanovski polynomials' was recently suggested in [47, 50] though under the wrong presumption that they were originally discovered in Routh' work.

The main purpose of this paper is to study the quantization scheme for $i$-GRef potentials in a more systematic fashion. The paper is structured as follows. In Section 2 we introduce manifolds of polynomial fraction (PFr) beams with two finite singular points $-\iota$ and $\iota$ referred to as $r$-GRef [1] and $i$-GRef [2] for $\iota = 1$ or $i$, respectively.

For each PFr beam there is a unique Liouville transformation $x \to \iota\eta$ which converts the Schrödinger equation into the differential equation of hypergeometric type [42, 43].

Section 3 links Stevenson's quantization technique of the differential equation with two complex-conjugated singular points to more conventional solutions in terms of Jacobi polynomials with complex conjugated indexes [23, 25, 31, 32] or, to be more precise, in terms of Romanovski-Routh polynomials.

In Section 4 the outlined technique is used to derive the quantization condition and bound eigenfunctions for the generic $i$-GRef potential. Starting from Section 5 we restrict our analysis solely to the Milson potential [2] generated using a symmetric change of variable. The most important result of this Section is the proof that each bound energy state in the Milson potential is determined by a positive root of a quartic equations. This very specific feature of the Milson potential allowed us to conclude that the m[th] eigenfunction formed by a Routh polynomial of order m is accompanied by the so-called [51-53] 'almost-everywhere-holomorphic' (AEH) solution which can be used as the factorization function (FF) to construct a new rational potential 'conditionally exactly'[x) quantized by polynomials (P-CEQ) as far as it does not have nodes inside the quantization interval. We refer to the latter polynomials as 'Routh-seed' ($\Re$S) Heine polynomials, by analogy with the term 'Gauss-seed' (GS) Heine polynomials used by us [51-53] for rational SUSY partners of the $r$- and $c$-GRef potential solvable by hypergeometric and confluent hypergeometric functions.

It was revealed that the Milson potential has two branches referred to as 'inside' and 'outside' depending on positions of zeros of the so-called 'tangent polynomial' (TP) relative to the unit circle. The two intersect along the shape-invariant Gendenshtein (Scarf II) potential. Based on the general theorem sketched in Appendix C for arbitrary invariant-under-reflection irregular solutions of the Schrödinger equation with a symmetric potential we proved that the RCSLE associated with the inner branch of the Milson potential has two infinite sequences of nodeless AEH solutions which can be used as FFs to construct new P-CEQ potentials with inserted ground-energy bound states. This assertion also remains valid for the border case represented by the Gendenshtein potential. In Section 6 we illuminate this limiting case in more details. Recently Quesne [49] ran into solutions of this type for the Gendenshtein potential and theorized that some of them may be nodeless and therefore can be used for constructing new potentials quantized by orthogonal polynomials of novel type. The proof presented in Section 6 fully confirmed

_______________________

[x) We use Junker and Roy's [54, 55] term 'conditionally exactly' to stress that the positions of singular points of the resultant potential depend on other parameters.

Quesne's conjecture.

The paper is concluded with a brief outline of some future developments accompanied by the aforementioned appendices: Appendix A analyzing necessary prerequisites for exact solvability of the Gendenshtein potential based on its shape-invariance, Appendix B presenting the precise definition of Routh polynomials, and Appendix C with the general proof that any invariant-under-reflection irregular solution of the 1D Schrödinger equation with an arbitrary symmetric potential is nodeless as far as it lies below the ground energy level.

## 2. Beams of Gauss Algebraic Fractions and Related Solvable Rational Potentials

In following Bose [4], we start from the rational SL equation written in the canonic form (RCSLE)

$$\left\{ \frac{d^2}{d\eta^2} + I[\eta; \varepsilon \mid {}_{\pm\iota}\boldsymbol{G}] \right\} \Phi[\eta; \varepsilon \mid {}_{\pm\iota}\boldsymbol{G}] = 0 \tag{2.1}$$

which has three regular singular points $\eta = -\iota, \; +\iota,$ and $\infty$. To be more precise, the Bose invariant of our current interest (using Milson's terms [2]) is defined as follows

$$I[\eta; \varepsilon \mid {}_{\iota}\boldsymbol{G}] \equiv I^{O}[\eta; -\iota, \iota; {}_{\iota}\bar{h}_{o}, {}_{\iota}O_{0}^{o}] + \wp[\eta; -\iota, \iota] \; \varepsilon, \tag{2.2}$$

where the so-called 'reference polynomial fraction' (RefPFr) and the nonnegative density function have the form

$$I^{O}[\eta; -\iota, \iota; {}_{\iota}\bar{h}_{o}, {}_{\iota}O_{0}^{o}] \; = -\frac{1}{4}[h_{o;-\iota}/(\eta+\iota)^2 + h_{o;\iota}/(\eta-\iota)^2 - {}_{\iota}O_{0}^{o}/(\eta^2-\iota^2)] \tag{2.3}$$

and

$$\wp[\eta; -\iota, \iota] = T_{K}[\eta; -\iota, \iota] \, /(\eta^2-\iota^2)^2, \tag{2.4}$$

with the $T_{K}[\eta; -\iota, \iota]$ standing for the TP of the $K^{th}$ order with no zeros between the end points of the quantization interval. In particular, this requirement implies that the polynomial $T_{K}[\eta; -i, i]$ must have a negative discriminant, as already noticed by Milson [2].

In this paper we focus solely on an analysis of the generic $i$-GRef potential generated using the second-order TP (K=2) so that the t-Schrödinger potential (K=0) -- 'trigonometric Rosen-Morse' potential in the classification scheme of Cooper et al [19, 20] -- is not covered by the current discussion. The parameters $_1h_{o;-1}$, $_1h_{o;1}$, and $_iO_0^o$ for $\iota = 1$ or $i$ are assumed to be real; the real parameter $_iO_0^o$ is supplemented by the complex-conjugated parameters

$$h_{o;i} \equiv h_o \equiv h_{o;R} + ih_{o;I}, \ h_{o;-i} \equiv h_o^* \equiv h_{o;R} - ih_{o;I} \qquad (2.5)$$

for $\iota = i$.

Though our initial technique [1] dealing with hypergeometric functions of the variable $z(x)$ varying between 0 and 1 is necessary to prove that the given potential is indeed exactly solvable (as well as to derive close-form expressions for the scattering amplitudes [56, 57]), use of the variable $_1\eta(x) = 2z(x) - 1$ suggested by Levai [31, 58] allows one to treat $r$- and $i$-GRef potentials in a symmetric fashion (with no drawback as far as we are only interested in bound state solutions). It is especially convenient for an analysis of the symmetric reduction of the $r$-GRef potential which will be discussed in detail in a separate publication [59]). Keeping in mind that the RCSLEs associated with the variables z(x) and $\eta(x) = 2z(x) - 1$ have different lower singular points ($e_0 = 0$ and $e_0 = -1$, accordingly) we use a slightly different notation for the appropriate PFr beams $\mathbf{B} = {}_1\boldsymbol{G}$ and $\mathbf{B} = {}_{\pm 1}\boldsymbol{G}$, despite the fact that they correspond to the same rational Liouville potential (RLP)

$$V[z(x) \mid {}_1\boldsymbol{G}] \equiv V[\eta(x) \mid {}_{\pm 1}\boldsymbol{G}]. \qquad (2.6)$$

To distinguish between the appropriate TPs we explicitly specify the quantization interval of the appropriate RCSLE, i. e.,

$$T_2[2z - 1; \pm 1] = 4T_2[z; 0, 1] \qquad (2.7)$$

On other hand, the notation $\mathbf{B} = {}_i\boldsymbol{G}$ and $\mathbf{B} = {}_{\pm i}\boldsymbol{G}$ is used by us in a completely interchangeable way.

It is worth stressing in this connection that we are interested only in GRef potentials which can be exactly quantized (not just solvable) by polynomials. Contrary to the statements made in [58, 60, 61], none of the $r$-GRef potentials generated using the TPs $T_1[\eta; -1,1] = \frac{1}{2}b_0\eta$, $T_2[\eta; -1,1] = \frac{1}{4}a\eta^2$, and $T_2[\eta; -1,1] = \frac{1}{4}a\eta(\eta+1)$, respectively, can be analytically quantized via Jacobi polynomials since the eigenfunctions discussed in the cited works do not vanish at $\eta=0$ where all three potentials have a singularity.

In following Milson's suggestion [2], we refer to PFr (2.2) as the 'Bose invariant'. In our terms the Bose invariant defines a GRef PFr beam $_\iota\mathcal{G}$ generated using a fixed TP $T_K[\eta; -\iota, \iota]$. As pointed by Milson [2], the Liouville transformation [5-9] of the RCSLE leads to the Schrödinger equation with the potential [4]

$$V[_\iota\eta(x) \mid _\iota\mathcal{G}] = -\wp^{-1}[_\iota\eta(x); -\iota, \iota]] \, I^O[_\iota\eta(x); -\iota, \iota; _\iota\bar{h}_o, _\iota O_0^o] - \frac{1}{2}\{_\iota\eta, x\}, \quad (2.6)$$

where the $\varepsilon$-independent change of variable $_\iota\eta(x)$ satisfies the first-order differential equation

$$_\iota\eta' = \wp^{-1/2}[_\iota\eta; -\iota, \iota] \quad (2.7)$$

(with prime standing for the derivative with respect to x) and $\{_\iota\eta, x\}$ is the so-called 'Schwartz derivative' [62]. (The appropriate historical bibliography can be found in the recent review [3]; however, we found it more convenient to refer the reader to McIntosh' textbook [61] which contains a complete list of the all formulas necessary for our analysis.)

It is convenient to parameterize the TP $T_K[\eta; -\iota, \iota]$ in the right-hand side of (2.4) as

$$T_K[\eta; -\iota, \iota] = [c_{-\iota}(\eta - \iota)^2 + c_\iota(\eta + \iota)^2 + _\iota d(\eta^2 - \iota^2)]/4 \quad (2.8)$$

with the leading coefficient

$$_\iota a = \frac{1}{4}(c_{-\iota} + c_\iota + _\iota d), \quad (2.9)$$

(assuming that $c_{-i}^* = c_i \equiv c_R + ic_I$ for $\iota = i$ and $_\iota d$ is real in both cases: $\iota = 1$ and $\iota = i$).

As pointed by Roychoudhury et al [63] in connection with the $r$-GRef potential, one needs to distinguish between the parameters $_\iota h_{o;-\iota}$, $_\iota h_{o;\iota}$, and $_\iota O_0^o$ in numerators of the fractions in the right-hand side of (2.3) and the coefficients $c_\iota$, $c_{-\iota}$, $_\iota d$ of TP (2.8). The latter are used to specify the change of variable, $_\iota\eta(x)$, which is chosen to be the same for all the PFr 'rays' are obtained by varying 'ray identifiers' $_\iota h_{o;-\iota}$, $_\iota h_{o;\iota}$, and $_\iota O_0^o$ within allowed limits. In other words, each PFr ray from the given "PFr beam" $_\iota\boldsymbol{G}$ represents the particular potential curve on potential surface (2.6).

The Bose invariant of our interest takes the form:

$$I[\eta;\varepsilon \mid {}_\iota\boldsymbol{G}] = -\frac{1}{4}\left[\frac{h_\iota(\varepsilon \mid {}_\iota\boldsymbol{G})}{(\eta-\iota)^2} + \frac{h_{-\iota}(\varepsilon \mid {}_\iota\boldsymbol{G})}{(\eta+\iota)^2} - \frac{O_0(\varepsilon \mid {}_\iota\boldsymbol{G})}{\eta^2-\iota^2}\right], \qquad (2.10)$$

where

$$h_{\pm\iota}(\varepsilon \mid {}_\iota\boldsymbol{G}) \equiv h_{o;\pm\iota} - c_{\pm\iota}\varepsilon \ , \qquad (2.11a)$$

and

$$O_0(\varepsilon \mid {}_\iota\boldsymbol{G}) \equiv {}_\iota O_0^o + {}_\iota d\ \varepsilon . \qquad (2.11b)$$

Note that each of the four shape-invariant potentials listed in [1] is in one-to-one correspondence with the particular $r$-GRef beam generated using the TP $T_K[\eta;\pm 1]$ with no roots other than $-1$ and $1$. Similarly, the t-Schrödinger and Gendenshtein potentials mentioned in the Introduction: are in one-to-one correspondence with the $i$-GRef beams generated using the TPs $T_0[\eta;-i,i] = c_0$ and $T_2[\eta;-i,i] = {}_i a\,(\eta^2+1)$, respectively.

In this paper we pay a special attention to the particular case of the $i$-GRef beam generated by means of the symmetric TP (sym-TP)

$$_iT_{sym}[\eta] = {}_i a\ (\eta^2 + \kappa_+). \qquad (2.12)$$

It is obtained (2.8) by choosing the coefficient $c_i$ to be real:

$$c_{+i} = c_{-i} \equiv c \qquad (2.12*)$$

and then replacing it for the new parameter

$$\kappa_+ = 1 - c/\,{}_i a. \qquad (2.12')$$

The appropriate potential has been studied in detail by Milson [2] and will be referred to as the Milson potential $V[\eta | \mathbf{M}]$. The remarkable feature of this potential (already pointed by Milson [2]) is that it intersects with sym-TP $r$-GRef potential on the surface formed by symmetric potential curves. The commonly cited symmetric Ginocchio potential [64] presents a typical example of such curves.

The common most important feature of RCSLEs (2.1) with Bose invariants (2.10) with $\iota = 1$ and $\iota = i$ is that it can be exactly quantized by polynomials. For the $r$-GRef potential both discrete energy levels and the appropriate eigenfunctions have been obtained in [1] based on the known asymptotics of the hypergeometric function of the real variable z at the singular point z =1. For $i$-GRef potentials the exact quantization of (2.1) presents a more challenging problem which was simply by-passed in [2]. As mentioned in Introduction, a very elegant technique for solving this SL problem has been developed in the nearly forgotten paper by Stevenson [12] who found bound state solutions of the mentioned equation by analyzing the asymptotic behavior of the hypergeometric function of complex variable

$$\xi_S = 2/(i\,\eta + 1) \quad \text{for } -\infty < \eta < \infty. \tag{2.13}$$

In next Section we present a detailed analysis of Stevenson's solution and its relation to the quantization technique in terms of orthogonal Romanovski-Routh polynomials.

## 3. Stevenson's Study on Bound Solutions of Differential Equation With Two Complex Conjugated Singular Points

The purpose of this Section is to rectify Stevenson's analysis [12] of bound state solutions of RSL equation (2.1) by directly linking his hypergeometric polynomials in complex variable (2.13) to a more familiar expression of the function $\Phi[\eta; \varepsilon |_i \mathbf{G}]$ in terms of a Jacobi polynomial with complex conjugated indexes [23, 25, 29, 31, 32, 37] or, to be more precise, to a set of mutually orthogonal real functions expressible in terms of Romanovski-Routh polynomials [30]. Compared with the conventional approach dealing with solutions of the Jacobi equation on the imaginary axis, the main advantage of Stevenson's analysis is the accurate proof that RSL equation (2.1) does not have any normalizable solutions other than those expressible in terms of a final set of hypergeometric polynomials in a complex argument. It has been pointed by Stevenson that these polynomials could be easily converted to real polynomials in a real argument. As it is explicitly shown below, the latter polynomials turn out to be nothing but Romanovski-Routh polynomials.

To directly relate our analysis to [12], first note that the substitution

$$\lambda + 2i\mu = h, \tag{3.1a}$$

$$4l(l+1) = 2Re\,h - O_0, \tag{3.1b}$$

with h, h*, and $O_0$ standing for parameters (2.11a) and (2.11b) in previous Section, converts Stevenson's Eq. (2) into (2.1) above, with $\iota$ set to $i$. To construct a solution regular at infinity, Stevenson introduced new complex variable (2.13) via the linear fractional transformation

$$1 - i\eta = 2(\xi_S - 1)/\xi_S \tag{3.2}$$

which converted singular points -i, i, and $\infty$ into 1, $\infty$, and 0. Since the variables $\eta$ and $\xi_S$ are related via the linear fractional transformation, the Schwartz derivative $\{\eta, \xi_S\}$ vanishes [62] so that RCSLE (2.1) for the function $\Phi[\eta; -i, i; \varepsilon \mid {}_i\mathcal{G}]$ takes the form:

$$\left\{\frac{d^2}{d\xi_S^2} + I_S[\xi_S]\right\}\Phi_S[\xi_S] = 0, \tag{3.3}$$

where

$$I_S[\xi] = -\frac{h*}{4\xi^2} - \frac{h}{4\xi^2(\xi-1)^2} - \frac{O_0}{4\xi^2(\xi-1)}. \tag{3.4}$$

Setting

$$h \equiv 4\rho(\rho - 1) \tag{3.5}$$

and making use of (3.1b) thus gives

$$I_S[\xi] = -\frac{\rho(\rho-1)}{(1-\xi)^2} - \frac{l(l+1)}{\xi^2} + \frac{\rho*(\rho*-1) - l(l+1) - \rho(\rho-1)}{\xi(1-\xi)}, \tag{3.6}$$

By choosing $\lambda_0 = 2l + 1$, $\lambda_1 = 2\rho - 1$, $\mu = 1 - 2\rho*$ in (5) in [1], we come to the following solution:

$$\Phi_S[\xi_S] = \xi_S^{l+1}(1 - \xi_S)^\rho F(\rho + \rho* + l, \ \rho - \rho* + l + 1, 2l + 2; \xi_S) \qquad (3.7)$$

regular at the singular point $\xi_S = 0$ for $l > -1$.

Taking into account that the functions $\Phi[\eta; \varepsilon \mid {}_i\boldsymbol{G}]$ and $\Phi_S[\xi_S]$ are related via the square root of the first derivative of $\xi_S$ with respect to $\eta$, one finds

$$\Phi[2i / \xi_S - 3i; \varepsilon \mid {}_i\boldsymbol{G}] = b_S \ \Phi_S[\xi_S] / \xi_S, \qquad (3.8)$$

where the scale factor $b$ will be defined below. We thus come to Stevenson's expression for the solution regular at the infinity:

$$\Phi[\eta; \varepsilon \mid {}_i\boldsymbol{G}] = b_S (1 + i\eta)^{-l-\rho}(1 - i\eta)^\rho F[\rho + \rho* + l, \ \rho - \rho* + l + 1, 2l + 2; 2/(1 + i\eta)] \qquad (3.9)$$

Note that the base of each exponent in the right-hand side of (3.9) is chosen in such a way that its real part is a positive constant so that the exponent phase is an unambiguous function of $\eta$. As proven by Stevenson, the solution remains finite at the points $\eta = \pm 3$ on the circle $|\eta - i| = 2$ only if the hypergeometric series terminates, i.e., iff

$$-2 \operatorname{Re} \rho - l = n \geq 0, \qquad (3.10)$$

where n is a nonnegative integer. By introducing a new complex parameter $\lambda \equiv 2\rho - 1$ we can alternatively represent the latter condition as

$$l = -\operatorname{Re} \lambda - n - 1. \qquad (3.10*)$$

Taking into account that the second term in (15.3.7) in [28] vanishes for $a = -n$, one finds

$$F[-n, \ b, c; \xi] = (-\xi)^{-n} \frac{\Gamma(c)\Gamma(b + n)}{\Gamma(b)\Gamma(c + n)} F[-n, 1 - c - n, \ 1 - b - n; 1 / \xi]. \qquad (3.11)$$

Writing (5.14.6) in [28] as

$$F[-n, \lambda + \lambda * + n + 1, \lambda * + 1; z] = \frac{(-1)^n n!}{(\lambda * + 1)_n} P_n^{(\lambda, \lambda *)}(2z - 1) \tag{3.12}$$

and setting $1 - c \equiv -\lambda *$, $1 - b \equiv -2Re\lambda - n$ in (3.11), one finds

$$\xi_S^n F[-n, -\lambda * - n, -2(Re\lambda + n); \xi_S] = \frac{(-1)^n n!}{(\lambda * + 1)_n} P_n^{(\lambda, \lambda *)}(i\eta), \tag{3.13}$$

By choosing

$$b_S = (i)^n \frac{(\lambda * + 1)_n}{n!}, \tag{3.14}$$

and setting $\lambda \equiv 2\rho - 1$ we can thus represent (3.9) as

$$\Phi[\eta; \varepsilon \mid {}_i \boldsymbol{G}] = (-i)^n \left| (1 - i\eta)^\rho \right|^2 \mathfrak{R}_m^{(\lambda *)}[\eta] \tag{3.15}$$

where $\mathfrak{R}_n^{(\alpha)}[\eta]$ is the Routh polynomial defined according to (B.15) in Appendix B.

This confirms that the right-hand side of (3.9) is indeed a real function of $\eta$, in agreement with Stevenson's assertion made in a similar context. Note that function (3.7) is a solution of differential equation (3.4) for any value of $l$ in (3.10*); however, it is regular at infinity only if $l > -1/2$. Substituting the latter constraint into (3.10*) thus gives

$$-Re\lambda > n + \frac{1}{2}. \tag{3.16}$$

It then directly follows from the analysis presented in Appendix B that the Routh polynomial $\mathfrak{R}_n^{(-\lambda *)}[\eta]$ turns into a Romanovski-Routh polynomial of the same order n iff rational part of its complex index satisfies condition (3.16).

## 4. Quantization condition for the generic *i*-GRef potential

By making use of (2.3) and (2.6) with $\iota = i$, the generic *i*-GRef potential can be represented as

$$V[\eta; -i, i \mid {}_i\boldsymbol{G}] = \frac{2h_{o;R}(\eta^2 - 1) - {}_iO_0^0(\eta^2 + 1) + 4h_{o;I}\,\eta}{4\,T_K[\eta; -i, i]} - \frac{1}{2}\{\eta, x\}\,. \tag{4.1}$$

Here we are interested only in TPs with negative discriminant so that the change of variable $x(\eta)$ is reversible at any $\eta$ between $-\infty$ and $+\infty$. There are only two cases when the TP does not have zeros other than singular points $-i$ and $i$, namely, 1) ${}_i a = 0$, $c_i = c_{-i}^*$ and 2) $c_i = c_{-i} = 0$, which correspond to the two shape-invariant potentials mentioned in Introduction.

For ${}_i a \neq 0$ it directly follows from the definition of the Schwartz derivative [62]:

$$\{\eta, x\} = \left(\eta''/\eta'\right)' - \frac{1}{2}\left(\eta''/\eta'\right)^2, \tag{4.2}$$

that

$${}_i a \lim_{|x| \to \infty} \{\eta, x\} = -\frac{1}{2} \tag{4.3}$$

and therefore (keeping in mind that ${}_i a > 0$) the appropriate PFr potential

$$V[\eta; -i, i \mid {}_i\boldsymbol{G}] = -\frac{4h_{o;R} - 4h_{o;I}\eta + \eta^2 + 1}{4T_2[\eta; -i, i]} - \frac{1}{2}\{\eta, x\} \tag{4.4}$$

vanishes at both limits $\eta \to -\infty$ and $\eta \to +\infty$, iff

$${}_iO_0^0 = 2h_{o;R} + 1. \tag{4.5}$$

or, making use of (3.1b), (2.11a), and (2.11b), iff

$$(2l + 1)^2 = -({}_i d + 2c_R)\varepsilon, \tag{4.6}$$

where $c_R \equiv Re\ c_i$. Substituting (2.9) and (3.10*) into the right- and left-hand sides of (4.6), respectively, thus leads to the following quantization condition:

$${}_i a\ \varepsilon_n = -[\lambda_{-;R}(\varepsilon_n) + n + 1/2]^2\,, \tag{4.7}$$

where $\lambda_{-;R}(\varepsilon)$ is the real part of the square root

$$\lambda_-(\varepsilon) \equiv \lambda_{-;R}(\varepsilon) + i\,\lambda_{-;I}(\varepsilon) = \sqrt{h_o + 1 - c_R\,\varepsilon} \qquad (4.8)$$

chosen via the requirement

$$\lambda_{-;R}(\varepsilon) < 0 \,. \qquad (4.8')$$

Representing (4.7) as

$$\sqrt{-_i a\,\varepsilon_n} + \lambda_{-;R}(\varepsilon_n) + \tfrac{1}{2} = -n \qquad (4.9)$$

one comes to the quantization condition found by Milson [2], with $\lambda_{-;R}(\varepsilon) + \tfrac{1}{2}$ here standing for Milson's parameter $\sigma$ (see next Section for details).

It is convenient to represent (4.8) in a slightly different form:

$$\lambda_-(\varepsilon \mid {}_i\boldsymbol{G}) \equiv \sqrt{{}_i\lambda_o^2 - c\varepsilon} \,, \qquad (4.10)$$

where the square root is chosen to satisfy the condition

$$Re\,\lambda_-(\varepsilon \mid {}_i\boldsymbol{G}) < 0 \qquad (4.10')$$

and

$$_i\lambda_o \equiv \lambda_{o;R} + i\,\lambda_{o;I} \equiv \sqrt{h_o + 1} \qquad (\lambda_{o;R} > 0,\ -\infty < \lambda_{o;I} < \infty). \qquad (4.11)$$

A solution expressible in terms of a Routh polynomial of order n exists at zero energy iff $\lambda_{o;R} = n + \tfrac{1}{2}$. As $\lambda_{o;R}$ slightly increases:

$$\delta\lambda_{o;R} \equiv \lambda_{o;R} - n - \tfrac{1}{2} << 1, \qquad (4.12)$$

one can approximate (4.10) at small $\varepsilon = \delta\varepsilon$ as

$$\lambda_-(\delta\varepsilon \mid {}_i\boldsymbol{G}) \approx -n - \tfrac{1}{2} - \delta\lambda_{o;R} - i\,\lambda_{o;I} - \frac{c\delta\varepsilon}{2(n + \tfrac{1}{2} + i\,\lambda_{o;I})} \qquad (4.13)$$

so that

$$\lambda_{-;R}(\delta\varepsilon \mid {}_i\boldsymbol{G}) \sim -n - \tfrac{1}{2} - \delta\lambda_{o;R}\,. \qquad (4.14)$$

and condition (4.7) holds at

$$_i a\,\delta\varepsilon \approx -(\delta\lambda_{o;R})^2 \,. \qquad (4.15)$$

Since $\lambda_{-;R}(\delta\varepsilon \mid {}_i\mathcal{G}) < -n - \frac{1}{2}$, the appropriate wavefunction, according to (3.16), describes a bound state. We thus conclude that that the number of bound levels in the $i$-GRef potential is given by the relation:

$$n_{max} = [\lambda_{o;R} - \tfrac{1}{2}]. \tag{4.16}$$

Making use of (3.15) we can represent the eigenfunctions as

$$\Phi[\eta;\varepsilon_n \mid {}_i\mathcal{G}] = (1-i\,\eta)^{\rho_n}(1+i\,\eta)^{\rho_n^*}\mathfrak{R}_m^{(\lambda_n^*)}[\eta], \tag{4.17}$$

where

$$\lambda_n = \lambda_{n;R} + i\lambda_{n;I} \equiv \lambda_-(\varepsilon_n) = 2\rho_n - 1. \tag{4.18}$$

By defining Romanovski-Routh polynomials $R_n^{(p,q)}(\eta)$ are defined via (32) in [37] and comparing the latter with (B.10) in Appendix B below we find

$$R_n^{(p,q)}(\eta) = \mathfrak{R}_n^{(-p+iq/2)}[\eta] \qquad \text{with } 0 \le n < \tfrac{1}{2}(p - \tfrac{1}{2}). \tag{4.19}$$

Note that the exponents $\rho_n$ depend on the energy $\varepsilon_n$ so that Romanovski-Routh polynomials in the right-hand side of (4.17) belong to different finite orthogonal sets, with the Gendenshtein potential as the only exclusion (see Section 6 for details).

## 5. Representing Quantization Condition for the Milson Potential as Real Quartic Equation

Using TP (2.12) with $\iota = i$ and ${}_ia = 1$ to generate the symmetric change of variable $\eta_+(x)$ via the first-order differential equation

$$\eta_+'(x) \equiv \frac{1+\eta_+^2(x)}{\sqrt{T_{sym}(\eta_+;-i,i)}}, \tag{5.1}$$

we come to the Milson potential [2]

$$V[\eta_+ \mid \mathbf{M}] = -\frac{4\,h_{o;R} - 4h_{o;I}\,\eta_+ + \eta_+^2 + 1}{4(\eta_+^2 + \kappa_+)} - \tfrac{1}{2}\{\eta_+, x\}\,,\qquad (5.2)$$

where [2]

$$\{\eta_+, x\} = \frac{1 - \eta_+^2}{\eta_+^2 + \kappa_+} - \left(\frac{1 + \eta_+^2}{\eta_+^2 + \kappa_+}\right)^2 \left[-\tfrac{1}{2} - \frac{\kappa_+ + 1}{\eta_+^2 + 1} + \frac{5\kappa_+}{2(\eta_+^2 + \kappa_+)}\right] \qquad (5.3)$$

and $\;h_O + 1 = \lambda_O^2 = -b + ic\;$ in Milson's notation.

Setting

$$\sigma = \tfrac{1}{2} + \lambda_{-;R}(\varepsilon)\,,\ \rho = \tfrac{1}{2} + i\lambda_{-;I}(\varepsilon) \qquad (5.4)$$

gives

$$\left(\sigma - \tfrac{1}{2}\right)^2 + \left(\rho - \tfrac{1}{2}\right)^2 = h_R(\varepsilon) + 1 \equiv h_{o;R} + 1 + (1 - \kappa_+)\varepsilon\,, \qquad (5.5a)$$

$$c = 2i(\rho - \tfrac{1}{2})(\sigma - \tfrac{1}{2}) = 2\lambda_{-;R}(\varepsilon)\lambda_{-;I}(\varepsilon) = Im\,h(\varepsilon) \equiv h_{o;I}\,. \qquad (5.5b)$$

[When deriving (5.5b), we also took into account that the coefficient of $\varepsilon$ in (2.11a),

$i\mathrm{c} = 1 - \kappa_+\,,$ is real.] Note that the appropriate formula for the parameter c in [2] differs

by the factor $(-1)$ from (5.5b).

Keeping in mind that

$$h_{o;R} + 1 = \lambda_{o;R}^2 - \lambda_{o;I}^2\,,\ \ |\,h_{o;R} + 1|^2 = \lambda_{o;R}^2 + \lambda_{o;I}^2\,, \qquad (5.6)$$

one can verify that Milson's condition for the total number of bound states in the sym-TP

$i$-GRef potential curve agrees with general formula (4.16) derived in previous Section.

As mentioned in Section 2 the latter curves form the intersection between $r$- and $i$-GRef

potentials. As a result it can be alternatively quantized in terms of both ultraspherical

[24] and Masjedjamei [65] (symmetric Romanovski-Routh) polynomials. We postpone

discussion of this remarkable feature of the sym-GRef potential and its SUSY partners for a separate publication [59].

Note that Milson came to the quantization condition $\sigma + \sqrt{-\varepsilon} = -\mathrm{n}$, starting from the

solution regular at $\eta_+ = i$:

$$\Phi[\eta_+;\varepsilon] = b_S \, (1 - i\eta_+)^{1/2 + \lambda_-(\varepsilon)/2} \, (1 + i\eta_+)^{1/2 + \lambda_-^*(\varepsilon)/2} \qquad (5.7)$$

$$\times F[1/2 + \lambda_{-;R}(\varepsilon) + \sqrt{-\varepsilon}, 1/2 + \lambda_{-;R}(\varepsilon) - \sqrt{-\varepsilon}; 1 + \lambda_-^*(\varepsilon); 1/2 + i\,\eta_+/2]$$

in our terms and noticing that it is normalizable if the hypergeometric series in the right-hand side truncates. This implicitly brought him to Jacobi polynomials with complex conjugated indexes and imaginary argument [see (B.14) in Appendix B below]. Though the resultant quantization condition turned out to be correct, the arguments used to prove it still leave open the question whether the given potential is exactly or only 'quasi-exactly' quantized. In fact, since the solution of our interest does not need to be regular at $\zeta = i$ and therefore one has to analyze an asymptotic behavior of a superposition of hypergeometric functions as it is done by Stevenson [12].

Representing (4.8) as

$$\lambda_{\pm;R}^2(\varepsilon) - \lambda_{\pm;I}^2(\varepsilon) = h_{o;R} + 1 - c\,\varepsilon, \qquad (5.8a)$$

$$2\lambda_{\pm;R}(\varepsilon)\,\lambda_{\pm;I}(\varepsilon) = h_{o;I}, \qquad (5.8b)$$

where $\lambda_+(\varepsilon)$ is square root with positive real part, and making use of (4.9), coupled with (2.12′), we conclude that $\lambda_{-;R}(\varepsilon_m)$ coincides with a positive root of the following quartic equation:

$$\lambda_{\dagger m;R}^4 - [h_{o;R} + 1 + (1 - \kappa_+)\,(\lambda_{\dagger m;R} + m + 1/2)^2]\lambda_{\dagger m;R}^2 - h_{o;I}^2 = 0 \qquad (5.9)$$

under the condition

$$\lambda_{\dagger m;R} < -m - 1/2 \quad (m = 0, 1, \ldots, n_{max}) \ \text{ for } \dagger = \mathbf{c}. \qquad (5.10)$$

(In following our 'obsolete' study [66] on Darboux transformations of radial potentials, we refer to functions regular and irregular at both end-points as solutions of types **c** and **d**, respectively.) Note that all the quartic equations share the same leading coefficient (regardless of values of the parameters $h_o$ and $\kappa_+ > 0$) and also the same negative free term except the limiting case of the symmetric potential curves ($h_{o;I} = 0$). Therefore each quartic equation must have a pair of real roots of opposite sign at least as far as the parameter $h_o$ remains complex ($h_{o;I} \neq 0$). (We shall come back to discussion of the limiting case $h_{o;I} = 0$ in the end of this Section.)

We thus conclude that there are two infinite sequences of Routh polynomials which are distinguished by sign of the root $\lambda_{\dagger m;R}$. Namely, the $m^{\text{th}}$ bound eigenfunction with the eigenvalue

$$_i\varepsilon_{\mathbf{c}m} \equiv \varepsilon_m = -(m + \tfrac{1}{2} - |\lambda_{\mathbf{c}m;R}|)^2 /_i a \quad (m = 0, 1, \ldots, n_{max}), \tag{5.11}$$

where $-m - \tfrac{1}{2} < \lambda_{\mathbf{c}m;R} < 0$, is accompanied by the AEH solution of type **d**,

$$\phi_{\mathbf{d}m}[\eta] = \sqrt{\eta^2 + 1}\ (1 - i\eta)^{\tfrac{1}{2}\lambda_{\mathbf{d}m}}(1 + i\eta)^{\tfrac{1}{2}\lambda^*_{\mathbf{d}m}}\mathfrak{R}_m^{(\lambda^*_{\mathbf{d}m})}[\eta] \tag{5.12}$$

defined by the root

$$\lambda_{\mathbf{d}m} = \sqrt{h_o + 1 - c\varepsilon_{\mathbf{d}m}} \qquad (m = 0, 1, \ldots) \tag{5.13}$$

with a positive real part $\lambda_{\mathbf{d}m;R} > 0$ so that the factorization energies

$$_i\varepsilon_{\mathbf{d}m} = -(\lambda_{\mathbf{d}m;R} + m + \tfrac{1}{2})^2 /_i a <\ _i\varepsilon_{\mathbf{c}m} \tag{5.14}$$

monotonically decrease with m. In our works [51-53] we use the term 'primary' for any sequence of polynomial solutions starting from a constant, i.e., each *primary* sequence of Routh-seed ($\mathfrak{R}S$) solutions by definition begins with the necessarily nodeless 'basic' solution

$$\phi[\eta_+; \dagger 0] = \sqrt{\eta_+^2 + 1}\ (1 - i\eta_+)^{\tfrac{1}{2}\lambda_{\dagger 0}}(1 + i\eta_+)^{\tfrac{1}{2}\lambda^*_{\dagger 0}/2}, \tag{5.15}$$

with $\dagger = \mathbf{c}$ or $\mathbf{d}$ in the particular case of the Milson potential. Each such solution can be used to construct the FF for the Darboux transformation (DT) either erasing ($\dagger = \mathbf{c}$) or inserting ($\dagger = \mathbf{d}$) the ground energy state. It has been proven in [53] the resultant potentials are quantized in terms of $\Re$S Heine polynomials which satisfy the second-order differential equation with two pairs of complex conjugated (purely imaginary) regular singular points $\pm i$ and $\pm\sqrt{\kappa_+}\, i$.

The second infinite sequence $\mathbf{d}'m$ of Routh polynomials forming irregular-at-both-ends $\Re$S solutions starts from a Routh polynomial of order $n_{max}+1$, where we use prime to indicate that we deal with the secondary sequence of seed solutions.

One can easily verify that

$$\lim_{m\to+\infty}\left[\, m\,|\,\lambda_{\dagger m; R}\,|\,\right] = |\,\hbar_{O; I}\,|\,/\sqrt{\kappa_+ -1} \qquad \text{for } \kappa_+ > 1 \qquad (5.16o)$$

and

$$\lim_{m\to+\infty}\left[\,\lambda_{\dagger m; R}\,/\,m\,\right] = l_\dagger \ \ \text{for } \kappa_+ < 1, \qquad (5.16i)$$

where

$$l_\mathbf{d} = \frac{\sqrt{1-\kappa_+}}{1-\sqrt{1-\kappa_+}} > 0 \qquad (5.16i\text{-d})$$

and

$$-1 < l_{\mathbf{d}'} = -\frac{\sqrt{1-\kappa_+}}{1+\sqrt{1-\kappa_+}} < 0. \qquad (5.16i\text{-d}')$$

We thus have to distinguish between 'inside' ($\kappa_+ < 1$) and 'outside' ($\kappa_+ > 1$) branches of the Milson potential depending on positions of TP zeros relative to the unit circle. Two branches intercept along the 'unit-circle' PFr beam $_i\mathbfcal{G}^{201}$ ($\kappa_+ = 1$) associated with the Gendenshtein potential. In this limiting case quartic equation (5.9) turns into the quadratic equation in $\lambda_R^2$ which is discussed in detail in next Section. The crucial difference between two branches comes from the fact magnitudes of both $\lambda_{\mathbf{d}m; R}$ and $\lambda_{\mathbf{d}'m;R}$ for $\kappa_+ < 1$ tend to $+\infty$ as m grows. This implies that there must be two infinite subsets of $\Re S$ solutions $\mathbf{d}m$ and $\mathbf{d}'m$ lying below the ground energy level.

In the limiting case of symmetric potential curves $V[\eta_+(x) | \mathbf{M}^{sym}]$ obtained by setting $h_{o;I}$ to 0 we can alternatively represent the appropriate quartic equation as

$$\kappa_+ \nu_{\mathbf{t}m}^2 - (2m+1)\nu_{\mathbf{t}m} + (m + \tfrac{1}{2})^2 - {}_i\lambda_o^2 = 0 \qquad (m = 0, 1, \ldots), \qquad (5.17)$$

where ${}_i\lambda_o$ is a nonnegative real number and

$$\nu_{\mathbf{t}m} \equiv \lambda_{\mathbf{t}m;R} + m + \tfrac{1}{2} \qquad\qquad (5.17a)$$

so that

$$\varepsilon_{\mathbf{t}m} = -\nu_{\mathbf{t}m}^2 / {}_i a. \qquad\qquad (5.17b)$$

Apparently quadratic equation (5.17) has a negative root only if ${}_i\lambda_o > \tfrac{1}{2}$. The number of excited bound energy states in the symmetric potential curve is thus equal to

$$n_{max} = [{}_i\lambda_o - \tfrac{1}{2}] \geq 0, \qquad\qquad (5.18)$$

in agreement with the general formula (4.16). An analysis of discriminant of quadratic equation (5.17),

$$\Delta_m^{sym}(h_{o;R}; \kappa_+) \equiv (1 - \kappa_+)(2m+1)^2 + \kappa_+(h_{o;R} + 1), \qquad\qquad (5.19)$$

shows that the RCSLE with the RefPFr $I^o[\eta_+ | \mathbf{M}^{sym}]$ has two infinite sequences of $\Re S$ solutions for $0 < \kappa_+ \leq 1$ and two finite sequences $\mathbf{t}m$ ($\mathbf{t} = \mathbf{d}, \mathbf{d}'$) with

$$2m < \frac{\kappa_+}{\kappa_+ - 1}(h_{o;R} + 1) - 1 \qquad\qquad (5.20)$$

if $\kappa_+ > 1$.

Eigenfunctions $\mathbf{c}m$ are described by negative roots of quadratic equation (5.17):

$$\kappa_+ \nu_{\mathbf{c}m} = m + \tfrac{1}{2} - \sqrt{(1 - \kappa_+)(m + \tfrac{1}{2})^2 + \kappa_+ {}_i\lambda_o^2} < 0 \quad \text{for } m = 0, \ldots, n_{max} \quad (5.21c)$$

whereas positive roots are given by either supplementary formula

$$\kappa_+ \nu_{\mathbf{dm}} = m + \tfrac{1}{2} + \sqrt{(1-\kappa_+)(m+\tfrac{1}{2})^2 + \kappa_{+i}\lambda_o^2} > 0 \qquad (5.21d)$$

or by extension of (5.21c) to larger values of m

$$\kappa_+ \nu_{\mathbf{d'm}} = m + \tfrac{1}{2} - \sqrt{(1-\kappa_+)(m+\tfrac{1}{2})^2 + \kappa_{+i}\lambda_o^2} \quad \text{for } m > n_{max}. \qquad (5.21d')$$

As expected $\nu_{\mathbf{d}0} > \nu_{\mathbf{c}0}$. Since the positive root $\nu_{\mathbf{dm}}$ monotonically increases with m we conclude that all the $\Re$S solutions $\mathbf{d}$m lie below the ground energy level $_i\varepsilon_{\mathbf{c}0}$ and therefore (as a direct consequence of the theorem proven in Appendix C) any $\Re$S solution $\mathbf{d}, 2\tilde{\mathbf{i}}$ formed by Routh polynomial of even order $2\tilde{\mathbf{i}}$ is necessarily nodeless.

If a pair of complex conjugated imaginary zeros of the TP lies either inside the unit circle $(0 < \kappa_+ < 1)$ or coincides with the singular points $\pm i$ $(\kappa_+ = 1)$ then $\nu_{\mathbf{d'm}}$ is proportional to $(1-\sqrt{1-\kappa_+})m/\kappa_+$ for sufficiently large m so that $\Re$S solutions $\mathbf{d'}, 2\tilde{\mathbf{i}}'$ lie below the ground energy level $\varepsilon_{\mathbf{c}0}$ and therefore may not have nodes as far as $\tilde{\mathbf{i}}'$ is chosen to be large enough. We thus conclude that there are two infinite sequences of nodeless $\Re$S solutions iff $\kappa_+$ does not exceed 1.

To be more precise, $\Re$S solutions $\mathbf{d'}, 2\tilde{\mathbf{i}}'$ for the inside branch of the symmetric potential are nodeless iff

$$\nu_{\mathbf{d'}, 2\tilde{\mathbf{i}}'} > |\nu_{\mathbf{c}0}|. \qquad (5.22a)$$

Making use of (5.21d') one can directly verify that the latter condition holds for

$$\tilde{\mathbf{i}}' > \frac{1}{\kappa_+}[\sqrt{\tfrac{1}{4} + \kappa_+({}_i\lambda_o^2 - \tfrac{1}{4})} - \tfrac{1}{2}]. \qquad (5.22b)$$

A criterion for existence of symmetric Routh polynomials of finite order with no zeros on real axis for the outside branch of the symmetric potential ($\kappa_+ > 1$) requires a more scrupulous examination postponed for a separate publication [59].

There are two general mechanisms which may result in formation of polynomial zeros on the real axis. The first is a merge of complex conjugated zeros into a double zero. The second is a reduction of the polynomial order when the leading coefficient vanishes.

The common feature of GS solutions of the RCSLEs associated with GRef potentials is that the polynomials forming these solutions (i.e., Jacobi, Laguerre, and Routh polynomials) may not have double zeros at regular points of the appropriate solved-by-polynomials equations. Otherwise both solution and its first derivative vanish which is only possible for zero solution.

Making use of (22.3.1) in [28] one finds that the leading coefficient of the Jacobi polynomial $P_m^{(\alpha^*,\alpha)}(y)$ with complex conjugated indexes depends solely on the common real part of its indexes:

$$P_{m;m}^{(\alpha^*,\alpha)} = \frac{(2\alpha_R + m)_m}{2^m m!},$$ (5.23)

with $(x)_m$ standing for the rising factorial. We conclude that the leading coefficient of any Routh polynomial from the series $\mathbf{d}m$ ($\alpha_R > 0$) must preserve its sign regardless of a value of its complex index.

The leading coefficient of any Routh polynomial from the sequence $\mathbf{d}'m'$ vanishes iff there is another polynomial solution of order $\tilde{m} < m'$ such that $\lambda_{\mathbf{d}'\tilde{m}'} = \lambda_{\mathbf{d}'m'}$ which can be true iff

$$2\lambda_{\mathbf{d}'m';R} + m' + \tilde{m}' + 1 = 0.$$ (5.24)

However, according to (5.16i-$\mathbf{d}'$),

$$-1 < 2l_{\mathbf{d}'} = -\frac{2\sqrt{1-\kappa_+}}{1+\sqrt{1-\kappa_+}} < 0$$ (5.25)

so that

$$\lim_{m' \to \infty} \frac{2\lambda_{\mathbf{d}'m';R} + m'}{m'} > 0.$$ (5.26)

Therefore the leading coefficient of a Routh polynomial of sufficiently large order from the secondary sequence may not vanish as imaginary part $h_{o;I}$ of the parameter $h_o$ varies at a fixed value of its real part.

We thus conclude that the RCSLE associated with the inside branch of the Milson potential has two infinite sets of nodeless $\Re S$ solutions $\mathbf{d},2\tilde{i}$ and $\mathbf{d}',2\tilde{i}'$ which can be used as FFs for constructing two new families of infinitely many P-CEQ potentials.

## 6. Gendenshtein potential as an illustrative example

One of by-side results of Gendenshtein's pioneering work [22] (usually ignored in the literature) is a brief mention of a new solvable potential

$$V_G(x) = \frac{b^2 - a(a+1) + b(2a+1)sh\,x}{2ch^2x} \tag{6.1}$$

commonly referred to as 'Scarf-II', in following the classification scheme of Cooper et al [19, 20]. Since analytical representation for solutions of the Schrödinger equation with potential (6.1) in terms of complexified Jacobi polynomials appeared later in the monograph [32] (possibly with no relation to the breakthrough paper of Dabrowska et al [25]) it is plausible that Gendenshtein has been aware of these solutions. However the author was unable to find any confirmation of this guess in the literature. Therefore, to stress the impact played by his work on studies of the mentioned complexified Jacobi polynomials by Dabrowska et al [25] we, in following Englefield and Quesne [21], refer to (6.1) as the Gendenshtein potential.

The crucial advantage of his parameters (A and B in [25] ), compared with Bagrov and Gitman's parameters $V_1$ and $V_2$ [32]:

$$4_i aV_1 = -4h_{o;R} - 3 = 1/4 - \lambda_{o;R}^2 + \lambda_{o;I}^2, \tag{6.2a}$$

$$4_i aV_2 = h_{o;I} = 2\lambda_{o;R}\lambda_{o;I} = (2a+1)\,b, \tag{6.2b}$$

is that they directly determine the square root of $h_o + 1$, namely,

$$_i\lambda_o \equiv {}_i\lambda_{o;R} + i\,{}_i\lambda_{o;I} \equiv \sqrt{h_o + 1} = a + \tfrac{1}{2} + i\,b \quad (_i\lambda_{o;R} > 0). \tag{6.3}$$

As a result, formulas (5.11) and (5.11*) for energies of AEH solutions take a simple form

$$_i\varepsilon_{\uparrow m} = -(m - a)^2 \qquad \text{for } \uparrow = \mathbf{c} \text{ or } \mathbf{d'} \tag{6.4}$$

and

$$_i\varepsilon_{\mathbf{d}m} = -(a + m + 1)^2. \tag{6.4*}$$

The remarkable feature of this limiting case is that exponent differences for the finite singular points $-i$ and $+i$ become energy-independent. Indeed, setting $\kappa_+ = 1$ turns quartic equations (5.9) into the quadratic equation in $_i\lambda^2_{R;\uparrow m}$,

$$_i\lambda^4_{R;\uparrow m} - (h_{o;R} + 1)\,_i\lambda^2_{R;\uparrow m} - h^2_{o;I} = 0, \tag{6.5}$$

with polynomial coefficients independent of m. Keeping in mind that the sought-for root $_i\lambda^2_{R;\uparrow m}$ of the latter equation must be positive one finds

$$_i\lambda_{\uparrow m} = -\,_i\lambda_{o;R} - \tfrac{1}{2} i\, h_{o;I} / \,_i\lambda_{o;R} \quad \text{for } \uparrow = \mathbf{c} \text{ or } \mathbf{d'} \tag{6.6}$$

$$_i\lambda_{\mathbf{d}m} = \,_i\lambda_{o;R} + \tfrac{1}{2} i\, h_{o;I} / \,_i\lambda_{o;R}, \tag{6.6*}$$

where

$$_i\lambda_{o;R} \equiv a + \tfrac{1}{2} = \sqrt{\tfrac{1}{2}(h_{o;R} + 1) + \sqrt{\tfrac{1}{4}(h_{o;R} + 1)^2 + h^2_{0;I}}}. \tag{6.7}$$

One can then verify that $-\,_i\lambda_{o;R}$ does coincides with the real part of the parameter $\nu$ in [32], as expected from Bagrov and Gitman's expression of eigenfunctions in terms of Jacobi polynomials in an imaginary argument. (The parameter $V_2$ in the above definitions must be divided by the factor $c^2$.)

Setting $\alpha = \lambda*$ in (B.15) in Appendix B and comparing the resultant expression with (3.15) we come to the expression for eigenfunctions given by Levai [31] in his Table 1. (Note that indexes of Jacobi polynomials in the appropriate expression in Table 1 of Dabrowska et al [25] are interchanged in error. The mentioned misprint in the expression of eigenfunctions in terms of Jacobi polynomials for the 'Scarf II' potential was later copied in [19, 20].)

Though Eqs. (11) in [38], with $_i\eta = -_i\lambda_o^*$ in our terms, match the expression for the eigenfunctions listed in Table 1 in [25], this is obviously an error caused by an accidental interchange of the indexes in the differential equation for the Jacobi polynomials, as one can easily verify by comparing (35) in [38] with (22.5.1) in [28].

As pointed to in [51], the energy independence of the exponent differences at two finite singular points assures that any SUSY partner of the Gendenshtein potential constructed using the nodeless FF

$$\phi[\eta_+ ; \varepsilon \mid {_i}\boldsymbol{G}^{201}; \boldsymbol{\dagger}m] = \left| (1 - i\eta_+)^{i\lambda \boldsymbol{\dagger} m + 1} \right| \mathfrak{R}_m^{(i\lambda \boldsymbol{\dagger} m)} [\eta_+] \qquad (6.8)$$

with a positive m and $\boldsymbol{\dagger} = \boldsymbol{d}$ or $\boldsymbol{d}'$ can be conditionally exactly quantized via a finite set of orthogonal $\mathfrak{R}S$ Heine polynomials, similar to those [51] constructed by Odake and Sasaki [68, 69] via the DTs of the hyperbolic Pöschl-Teller potential [67] using Romanovski-Jacobi polynomials [30] (in Lesky's terms [38]) to form the appropriate AEH FFs.

Recently the aforementioned AEH solutions of type $\boldsymbol{d}$ were discovered by Quesne [49] who speculated that there are even-order Routh polynomials ( 'Case III Romanovski polynomials' in her terms) which do not have real zeros and thereby can be used to generate new finite sets of orthogonal polynomials. This conjecture inspired the author to study this problem more carefully. The surprising (at least for the author) result was that no even-order Routh polynomial from series $\boldsymbol{d}$ as well as no Routh polynomial of even order $2\tilde{i}' > 2a$ from series $\boldsymbol{d}'$ has real roots. (In previous Section we proved a similar result for the inside branch of the Milson potential, except that we were unable to specify the starting polynomial order for the secondary subset formed by Routh polynomials with no real zeros.)

Indeed, it directly follows from (6.4) and (6.4*) that all $\mathfrak{R}S$ solutions $\boldsymbol{d}m$ lie below $_i\varepsilon_{\boldsymbol{c}0}$. This is also true for AEH solutions $\boldsymbol{d}'m'$ provided that $m' > 2a > 0$. Note that index $\alpha^*$ of $\mathfrak{R}_m^{(\alpha^*)}(\eta_+)$ becomes real if the asymmetry parameter b is equal to zero and as a result the Gendenshtein potential turns into the symmetric Rosen-Morse [70, 71] potential curves. Based on the analysis presented in Appendix C we conclude that all the AEH solutions $\mathbf{d}',2\tilde{i}'$ for $\tilde{i}' > a$ as well as any AEH solution $\mathbf{d},2\tilde{i}'$ are nodeless for $b=0$ and therefore Routh polynomials $\Re_{2\tilde{i}'}^{(-a-\frac{1}{2})}(\eta_+)$ for $\tilde{i}' > a > 0$ and $\Re_{2\tilde{i}}^{(a+\frac{1}{2})}(\eta_+)$ do not have zeros on the real axis. [The lower bound $2a$ for orders of Routh polynomials $\Re_{2\tilde{i}'}^{(-a-\frac{1}{2})}(\eta_+)$ with no real zeros is nothing but the condition (5.22b) in the limiting case $\kappa_+ = 1$.]

Taking into account that the parameter A in potential function (34) in [72] (or $\alpha$ in [73]) stands for $a + \frac{1}{2}$ we conclude that the functions $\phi_1(x)$ and $\phi_4(x)$ in [72] are nothing but the AEH solutions $\mathbf{d}'m$ and $\mathbf{d}m$ in our terms. In general the requirement that nodeless AEH solutions of this type must lie below the ground energy level [73] is necessary but not sufficient. However since the potential and selected AEH solution are even functions of x the imposed condition turned out to be both necessary and sufficient. The assertion that both solutions $\phi_1(x)$ and $\phi_4(x)$ are nodeless can be proven based on the Klein formula [74, 24] for the number of zeros of the appropriate Jacobi polynomials inside the interval $(-1, +1)$.

Coming back to the generic Gendenshtein potential, it directly follows from (5.23) that the order of a Routh polynomial may not be changed by varying an imaginary part of its index. This observation allows us to conclude that any Routh polynomial with no real zeros in the limit $\lambda_o = a + \frac{1}{2}$ preserves this feature for any values of the asymmetry parameter $b$. One can directly verify that leading coefficient (5.23) is positive for any Routh polynomial of order $2\tilde{i}' > 2a$ from the secondary sequence. We thus proved that any $\Re S$ solution $\mathbf{d},2\tilde{i}$ as well as all the solutions $\mathbf{d}',2\tilde{i}'$ for $\tilde{i}' > a$ are necessarily nodeless and therefore can be used for constructing P-CEQ SUSY partners of the Gendenshtein potential, as theorized by Quesne [49].

In following Quesne's [75] analysis of the quadratic cases for the Darboux-Poschl-Teller potential [27, 67] and the centrifugal oscillator let us consider second-order Routh polynomials as an illustrative example. Starting from the conventional expression for classical Jacobi polynomial of second order and extending the indexes to the field of complex numbers we come to the quadratic polynomial

$$\mathfrak{R}_2^{(\alpha)}(\eta) \equiv \mathfrak{R}_{2;2}(\alpha)\eta^2 + \mathfrak{R}_{2;1}(\alpha)\eta + \mathfrak{R}_{2;0}(\alpha) = -P_2^{(\alpha^*,\alpha)}(i\eta) \qquad (6.11)$$

with the real coefficients

$$\mathfrak{R}_{2;2}^{(\alpha_R + i\alpha_I)} \equiv \mathfrak{R}_{2;2}(\alpha_R + i\alpha_I) = -\tfrac{1}{4}(2\alpha_R + 3)(\alpha_R + 2), \qquad (6.12a)$$

$$\mathfrak{R}_{2;1}^{(\alpha_R + i\alpha_I)} \equiv \mathfrak{R}_{2;1}(\alpha_R + i\alpha_I) = -\tfrac{1}{2}\alpha_I(2\alpha_R + 3), \qquad (6.12b)$$

$$\mathfrak{R}_{2;0}^{(\alpha_R + i\alpha_I)} \equiv \mathfrak{R}_{2;0}(\alpha_R + i\alpha_I) = -\tfrac{1}{4}(2\alpha_I^2 + \alpha_R + 2) \qquad (6.12c)$$

and discriminant

$$\Delta_{\mathfrak{R}}(\alpha_R + i\alpha_I) = -\tfrac{1}{4}(2\alpha_R + 3)[(\alpha_R + 2)^2 + \alpha_I^2]. \qquad (6.13)$$

Since the Gendenshtein potential has at least three energy levels only if $_i\lambda_{o;R} > \tfrac{5}{2}$ $(a > 2)$ the Romanovski-Routh polynomial describing the second excited bound energy state $(\alpha_R = -_i\lambda_{o;R})$ has positive discriminant as expected. As expected, no second-order Routh polynomial has real roots for $\alpha_R < -\tfrac{3}{2}$. In case of the Routh polynomial $\mathfrak{R}_2^{(\alpha)}(\eta)$ from series **d′** ($\alpha_R = -_i\lambda_{o;R}$) the latter constraint is equivalent to the necessary and sufficient condition $2a < 2$ for the second-order Routh polynomial from this series not to have real roots on the real axis.

## 7.   Concluding Remarks and Future Developments

The main purpose of this paper was to incorporate the RCSLE with three regular singular points $-i$, $+i$, and $\infty$ into the general theory of PFr beams [53] either exactly or conditionally exactly quantized by GS Heine polynomials. The crucial reason for introducing this terminology is to separate the theory of PFr beams from their 1D quantum-mechanical realizations obtained via Liouville transformations. Some of the results for PFr beams may have merit on their own, with no reference to quantum mechanics. For instance, this is true for networks of GS Heine polynomials which can be introduced without even mentioning the RLP associated with the given RCSLE.

Though eigenfunctions of the Schrödinger equation with the generic $i$-GRef potential has been already analyzed by Milson [2] there are two crucial points which we found important to address. Firstly we incorporated Stevenson's theory [12] to prove that the potential is *exactly* quantized by Romanovski-Routh polynomials. The latter commonly referred to simply as Romanovski polynomials were introduced into quantum-mechanical applications of our interest by Kirchbach et al [35-37] years after publication of Milson's paper [2]. Secondly we proved that the energy spectrum of the potential generated using symmetric TP (2.12) is described by negative roots of a sequence of quartic equations. Since Milson's paper [2] mostly focused on a detailed analysis of this particular reduction of the generic $i$-GRef potential we refer to it as the Milson potential, $V[\eta_+ | \mathbf{M}]$.

The fact that the transcendental equation derived by us for the energy spectrum of the generic $i$-GRef potential turns into the quartic equation is fundamentally significant because it implies that the $m^{th}$ eigenfunction is always accompanied by an $\Re S$ solution $\mathbf{d}m$ irregular at infinity which (if nodeless!) can be used as the FF to construct a new potential quantized by the so-called '$\Re S$ Heine polynomials'.[x] (Since a pair of SUSY partners $V[\eta_+(x) | \mathbf{M}_{\downarrow\mathbf{d}0}]$ and $V[\eta_+(x) | {}^1\mathbf{M}_{\mathbf{d}0}]$ merge in the limiting case of the shape-invariant Gendenshtein potential, the quantized $\Re S$ Heine polynomials $Hi_n[\eta_+ | {}^1\mathbf{M}_{\mathbf{d}0}]$ associated with the RLP $V[\eta_+(x) | {}^1\mathbf{M}_{\mathbf{d}0}]$ turn into Romanovski-Routh polynomials if the coefficient $\kappa_+$ in (2.12) is set to 1.)

It has been proven that none of the Routh polynomials forming the primary sequence of $\Re S$ solutions $\mathbf{d}m$ has real roots either for the inside ($\kappa_+ < 1$) branch of the Milson potential or for the border case ($\kappa_+ = 1$) represented by the Gendenshtein potential. As far as $\kappa_+$

––––––––––––––––––

[x] In case of the generic $i$-GRef potential one can still construct its P-CEQ SUSY partners via either the single-step DT erasing the ground energy state or via double-step DTs [76-79] using a pair of sequential ('juxtaposed' [79]) eigenfunctions as seed solutions.

does not exceed 1 there is also another infinite subset of nodeless $\Re S$ solutions $\mathbf{d}'m'$ which can be used for constructing the second manifold of infinitely many P-CEQ potentials.

In the recent publication [49] Quesne came to the AEH solutions $\mathbf{d}m$ and $\mathbf{d}'m'$ for the Gendenshtein (Scarf II) potential from a different perspective and speculated that there are Routh polynomials (in our terms) with no zeros on the real axis. Her conjecture stimulated our renewed interest in this problem, compared with the first version of this paper. And we thus confirmed that there are two infinite sequences of nodeless even-order Routh polynomials which can be thus used for constructing infinitely many finite sets of orthogonal polynomials as suggested by Quesne.

It is worth emphasizing that our corroboration of Quesne's conjecture is essentially based on the general theorem proven in Appendix C below, namely, that any irregular invariant-under-reflection solution of the Schrödinger equation with a symmetric 1D potential is necessarily nodeless if it lies below the ground energy level. In particular this is true for any invariant-under-reflection AEH solution below the ground energy level. Therefore any SUSY partner of the symmetric quantized-by-polynomials potential generated using such a solution as the FF must be also quantized by polynomials.

The direct corollary of this finding is that the Wroskian of invariant-under-reflection seed solutions of the Schrödinger equation with an arbitrary symmetric potential does not have nodes inside the quantization interval if all the solutions lie below the ground energy level. This is the direct consequence of the fact that any solution of the Schrödinger equation with the multi-step SUSY partner of the original symmetric potential can be expressed as ratio of the appropriate Crum Wroskians [80] multiplied by a positive weight function. Since any such solution necessarily lies below the ground energy level and is itself an even function of x one can use mathematical induction to confirm that each Crum Wroskian in the given chain is nodeless as asserted.

In the specific case of the Hermite equation a set of irregular invariant-under-reflection AEH solutions at negative energies is formed by even Hermite functions of an imaginary argument. The fact that the latter functions do not have nodes on the real axis was explicitly exploited by Samsonov et al [81, 82,79] to construct rational SUSY partners of the harmonic oscillator which were quantized by orthogonal polynomials of a new type – the so-called 'X$_m$-Hermite' polynomials in Gomez-Ullate, Kamran, and Milson's [83] classification scheme of exceptional orthogonal polynomials. (The Samsonov-Ovcharov [81, 82] orthogonal polynomials were more recently re-dicovered by Fellows and Smith [84] in a very similar context.) Based on the arguments presented above we thus conclude that Crum Wroskians of even Hermite functions of an imaginary argument $ix$ may not have nodes at real values of x, in agreement with the accurate theory of multi-index exceptional Hermite polynomials developed by Gomez-Ullate, Grandati, and Milson [86] a short time ago.

It is worth to draw reader's attention to the fact that we use the term 'sym-GRef' without distinguishing between symmetric reductions of the $r$- and $i$-GRef potentials. As it has been already pointed by Milson [2], the symmetric $i$-GRef potential is nothing but another representation for the symmetric potential [1] obtained by reflection of the radial $r$-GRef potential $V[z|_1\mathcal{G}^{211}]$ in the particular case of zero centrifugal barrier (i.e., for the exponent difference at the origin, $\lambda_o$, to be set to $\frac{1}{2}$). (In particular, the Ginocchio potential [64] is obtained from the latter symmetric potential curves by simply setting the leading coefficient of the TP to a certain a priori selected value [86]). It will be proven in [59] that the reflected potential allows an alternative representation making use of the symmetric TP, with the RIs $_1h_0$ and $_1h_1$ set equal to each other. This implies that the sym-GRef potential curves form an intersection between the $r$-GRef and $i$-GRef potentials and therefore can be analytically quantized using both ultraspherical [24] polynomials and the definite-parity reduction of Romanovski-Routh polynomials referred to in our papers as Majedjamei [64] polynomials. We will exploit this finding in [59] by constructing the single-source net of symmetric rational potentials conditionally exactly quantized by ultraspehrical-seed ($\mathcal{U}$S) and Majedjamei-seed ($\mathcal{M}$S) Heine polynomials. In addition to irregular symmetric seed solutions the constructed Crum Wroskians will also include pairs of 'juxtaposed' eigenfunctions [75-78].

It should be stressed that one can extend nodeless definite-parity seed solutions to the 'nearly symmetric' Milson potential only by varying the parameter $h_{I;R}$ but not the linear coefficient of the TP (whose deviation from zero would also break the symmetry) because there is no certainty that the solution in question would retain the AEH form at nonzero values of this coefficient. Here we come to the important difference between the *i*- and *r*-GRef potentials, namely, energies of AEH solutions of the RCSLE associated with the generic *r*-GRef potential are described by the quartic equation [87, 88] for any values of potential parameters and therefore we can independently vary the TP linear coefficient while preserving the solution form.

A quick look at Quesne's rules [89] for Jacobi polynomials with negative Jacobi indexes not to have zeros between -1 and 1 shows that 'type-III' single-step SUSY partners of the RM potential exist only within a very artificial parameter range. This implies that construction of P-CEQ SUSY partners using irregular-at-both-end solutions is tricky even for single-step SUSY partners of the RM potential. Finding parameter ranges where Wroskians of seed solutions of type **d** are nodeless [90, 91] seems to be even a more challenging problem.

## Acknowledgement

I am expressing my special gratitude to Mariana Kirchbach for bringing my attention to Stevenson's paper as well as some valuable comments on the first draft of this paper. The mentioned reference helped me to bring this paper to the sufficiently high level of mathematical accuracy. It is my pleasure to thank Geza Levai for checking his old notes to confirm that indexes of Jacobi polynomials for the Scarf II potential were indeed interchanged in error by Dabrowska et al. I am also thankful to David Avarez-Castillo for double-checking the aforementioned result and to Robert Milson for some useful comments on a preliminary version of this paper.

## Appendix A

### Some Comments on Exact Solvability of the Gendenshtein potential

There is a wide-spread opinion that Gendenshtein did prove in his well-known paper [22] that any shape-invariant potential must be exactly solvable. The purpose of this Appendix is to demonstrate that Gendenshtein's arguments have to be accompanied by some additional assumptions which drastically reduce practical significance of his conclusions. The Gendenshtein potential was chosen as an example because we spent a lot of efforts in our earlier studies to prove that this potential is indeed exactly solvable as it was claimed in the literature. All those efforts lost their significance after the author became aware of Stevenson's paper [12]. However, it seems useful to reproduce at least the first part of our analysis to shed light on some pitfalls in Gendenshtein's arguments in support of the assertion that any shape-invariant potential is necessarily exactly solvable.

Suppose that RSL equation (2.1) associated with the Gendenshtein potential has a non-polynomial solution describing an 'exotic' bound state for $\lambda_{o;R} < -1/2$. Since the number of real nodes of each Romanovski-Routh polynomial coincides with the polynomial order, this energy level (if exists) must lie above those given by (6.4). It directly follows from (6.4) that the Darboux transformation eliminating the ground state increases the (originally negative) real part of the complex parameter (6.3) by 1 until it becomes larger than $-1/2$. Depending on sign of the asymmetry parameter $b$ potential (6.1) approaches zero from below at either $-\infty$ or $+\infty$ so that the resultant potential does have a well which might support a bound state.

To confirm that the Gendenshtein potential is exactly solvable, one needs to prove that the region $\lambda_{o;R} > -1/2$ does not support bound states. Before becoming aware of Stevenson's paper [12] the author had to put together a tricky proof based on to the Hellmann–Feynman theorem [92, 93] in support of this assertion. We do not need to reproduce it here since Stevenson's work provides a much more effective and general tool for solving this problem.

## Appendix B

## Routh Polynomials as Real Subset of Complexified Jacobi Polynomials in Imaginary Argument

Despite numerous citations to Routh' pioneering work [44] during last few years [36, 37, 46-50, 94], none of them referred it in a correct context. In particular, contrary to the statement by Weber [46] more recently repeated in [47, 48], Routh [44] did not consider polynomial solutions in complex plane. The focus of his study was the differential equation of hypergeometric type with real coefficients

$$(a\zeta^2 + b\zeta + c)\frac{d^2F}{d\zeta^2} + (f\zeta + g)\frac{dF}{d\zeta} + h\zeta = 0 \tag{B.1}$$

defined on the real axis. His main fundamental result is that this equation has a polynomial solution of order $m \geq 0$ iff

$$h = -m\,[f + a(m-1)]. \tag{B.2}$$

If the polynomial

$$A_2[\zeta] \equiv a\zeta^2 + b\zeta + c \tag{B.3}$$

has two distinct real roots then the polynomial solutions discovered by Routh turn into Jacobi polynomials with real indexes. In case of complex-conjugated singular points differential equation (B.1) can be represented as

$$(\zeta^2 + 1)\frac{d^2F}{d\zeta^2} + (f\zeta + g)\frac{dF}{d\zeta} + hF[\zeta] = 0 \tag{B.4}$$

which is the starting point of this analysis. It turns out [45] that Routh polynomials with $m < -f$ ($f < 0$) form an orthogonal subset discovered later by Romanovsky [30]. As mentioned in Introduction, we refer to them as Romanovski-Routh polynomials, by analogy with the terminology used by Lesky [38] for two other sets of orthogonal polynomials found by Romanovsky.

Taking into account that the characteristic exponents $\rho$ for the singular point $-i$ is given by one of complex roots of the quadratic equation

$$4\rho(\rho - 1) = h \equiv h_i(\varepsilon \,|\, {}_i\boldsymbol{\mathcal{G}}) \tag{B.5}$$

one can easily verify that the gauge transformation

$$\Phi[\eta; \varepsilon \,|\, {}_i\boldsymbol{\mathcal{G}}] = (1 - i\eta)^\rho (1 + i\eta)^{\rho^*}\, F[\eta; \varepsilon \,|\, {}_i\boldsymbol{\mathcal{G}}] \tag{B.6}$$

converts RCSLE (2.1) into the differential equation of hypergeometric type:

$$\left\{ (\eta^2 + 1)\frac{d^2}{d\eta^2} + 2[\rho(\eta - i) + \rho^*(\eta + i)]\frac{d}{d\eta} + C_0 \right\} F[\eta; \varepsilon \mid {}_i\mathcal{G}] = 0, \qquad (B.7)$$

where

$$C_0 = 2|\rho|^2 + \tfrac{1}{4} O_0(\varepsilon \mid {}_i\mathcal{G}). \qquad (B.8)$$

An analysis of the asymptotics of differential equation (B.5) at large $\eta$ shows that it may have a polynomial solution of order m only if

$$m(m-1) + 2m(\rho + \rho^*) + C_0 = 0, \qquad (B.9)$$

which is nothing but Eq. (6) in Routh' paper [44]. As proven by Routh, this is also the sufficient condition.

As pointed to by Ismail [45] (who inaccurately referred the assertion below to Routh [44]) Routh polynomials can be generated via the Rodrigues formula

$$\mathfrak{R}_m^{(\alpha)}(\eta) = \frac{1}{w^{(\alpha)}(\eta)}\frac{d^m}{d\eta^m}\left[(1 + \eta^2)^m w^{(\alpha)}(\eta)\right]. \qquad (B.10)$$

where the generating function

$$w^{(\alpha)}(\eta) = (1 + \eta^2)^{Re\,\alpha}\,exp\,(2\,Im\,\alpha\,atan\,\eta) \equiv (1 - i\eta)^\alpha(1 + i\eta)^{\alpha^*} \qquad (B.11)$$

satisfies the differential equation:

$$\frac{d}{d\eta}\left[(1 + \eta^2)\,w^{(\alpha)}(\eta)\right] = [(1 - \alpha)(\eta - i) + (1 - \alpha^*)(\eta + i)]\,w^{(\alpha)}(\eta). \qquad (B.12)$$

with $\alpha$ used instead of $a + ib$. For $Re\,\alpha < -1$ generating function (B.11) turns into the density function for a finite set of orthogonal Romanovski-Routh polynomials, with m in (B.10) restricted by the condition $2m+1 < -Re\,\alpha$.

If (B.9) is fulfilled, then the substitution $y = i\eta$ converts (B.7) into the Jacobi equation:

$$\left\{ (1 - y^2)\frac{d^2}{dy^2} + 2[\alpha_I - (\alpha_R + 1)\,y]\frac{d}{dy} + m(m + 2\alpha_R + 1) \right\} P_m^{(\alpha^*, \alpha)}(y) = 0, \quad (B.13)$$

with $\alpha \equiv 2\rho$-1. In following Kuijlaars et al [95], it is convenient to define Jacobi polynomials with complex parameters as

$$P_m^{(\beta,\alpha)}(y) = \frac{1}{2^m} \sum_{k=0}^{m} \frac{1}{k!(m-k)!}(\alpha)_k\,(\beta)_{m-k}\,(y-1)^k\,(y+1)^{m-k} \qquad (B.14)$$

with $(x)_n$ standing for the rising factorial. It is essential that the sum appearing in the right-hand side of (B.14) is a polynomial in each of three variables $y$, $\alpha$, $\beta$ and therefore all recurrence relations between its coefficients as well as between different polynomials of this type must hold when both the parameters $\alpha$ and $\beta$ and the argument $y$ are extended into the complex field. In particular, one can extend to complex indexes Askey's proof [29] that Jacobi polynomials $P_n^{(\beta,\alpha)}(i\eta)$ form a finite orthogonal set provided that $Re(\alpha+\beta) < -2n-1$. If $\beta = \alpha^*$, then these polynomials scaled via the relation

$$\mathfrak{R}_n^{(\alpha)}(\eta) = (-i)^n\,P_n^{(\alpha^*,\alpha)}(i\eta) \qquad (B.15)$$

remain real on the imaginary axis [36] [x)], forming an orthogonal set of real polynomials discovered by Romanovsky [30]. Using (22.5.42) in [28] with $x$ standing for $i\eta$ here, one can alternatively represent the Jacobi polynomials in the right hand-side of (B.15) as

$$P_n^{(\alpha^*,\alpha)}(i\eta) = \frac{(-1)^n(\alpha+1)_n}{n!}\,F[-n, n+1+2Re\,\alpha; 1+\alpha; \tfrac{1}{2}(1+i\eta)]. \qquad (B.16)$$

Complexified Jacobi polynomials (B.14) can be also constructed using the generating function:

$$w^{(\beta,\alpha)}(y) = (1-y)^\beta\,(1+y)^\alpha \quad \text{for } |Re\,y| < 1. \qquad (B.17)$$

---

[x)] Note that the factor $(1-x)^n$ in the definition (93) of Jacobi polynomials in [36] must be changed for $(x-1)^n$. The right-hand side of the cited equation describes the polynomial in $-ix$ which is real for both even and odd $n$. The right-hand side of (95) in [36] differs from the complex conjugate of this polynomial by the factor $(-1)^n$, which allowed the authors to come to the correct conclusion that the Jacobi polynomial $P_n^{(\alpha^*,\alpha)}(ix)$ of an odd order is imaginary for real $x$.

Comparing generating functions (B.11) and (B.17), one can directly verify that Routh polynomials defined via (B.10) are nothing but Jacobi polynomials (B.16) scaled via (B.15).

**Appendix C**

**SUSY ladders of symmetric potentials generated using (arbitrarily chosen) invariant-under-reflection solutions below the corresponding ground energy levels**

Both in this paper and in following publications we use nodeless irregular-at-both-ends AEH solutions of the RCSLE associated with the symmetric GRef potential as the starting point for constructing their nodeless extensions at nonzero values of the asymmetry parameters. The assertion that the solutions in question are nodeless in the limiting case of the symmetric RLP directly follows from the general feature of the 1D Schrödinger equation with a symmetric potential, namely, any invariant-under-reflection solution of this equation below the ground energy level may not have nodes inside the quantization interval.

In fact, let $\psi_{\mathbf{a}}(x;\varepsilon)$ be a regular-at-lower-end solution of the Schrödinger equation with a symmetric (invariant under reflection $x \to -x$) potential $V_{sym}(x)$. If it lies below the ground energy level $\varepsilon_{\mathbf{c}0}$ then it must be nodeless. (We refer the reader to Section 3 in [53] for some additional comments concerning applicability of the latter assertion to the infinite quantization interval.) Without loss of generality we can choose $\psi_{\mathbf{a}}(x;\varepsilon) > 0$. The positive function

$$\psi_{\mathbf{b}}(x;\varepsilon) = \psi_{\mathbf{a}}(-x;\varepsilon) > 0 \qquad\qquad (C.1b)$$

is thus a solution of the Schrödinger equation regular at the upper end so that the linear superposition of two solutions

$$\psi_{\mathbf{d}}(x;\varepsilon) = \psi_{\mathbf{a}}(x;\varepsilon) + \psi_{\mathbf{b}}(x;\varepsilon) > 0 \qquad\qquad (C.1d)$$

is an irregular-at-both-ends solution invariant under reflection $x \to -x$ and nodeless by definition.

In the particular case of the harmonic oscillator it was Sukumar [78] who explicitly proved that any invariant-under-reflection solution of the Schrödinger equation below the ground energy level is nodeless and therefore it can be used as the FF of the DT to construct an exactly quantized SUSY partner of the harmonic oscillator with an extra energy level inserted at the factorization energy.